\documentclass[a4paper,11pt]{article}

\usepackage{jinstpub} 
\usepackage{enumitem}
\usepackage{subcaption}
\usepackage{listings}
\usepackage{float}

\usepackage{newtxtext,newtxmath}
\title{SAGE: A Monte Carlo Simulation Framework for Experiments with Germanium Detectors}
\author[a]{Ze~She}
\author[a]{Hao~Ma}
\author[a]{Weihe~Zeng}
\author[a]{Wenhan~Dai}
\author[a]{Xinping~Geng}
\author[a]{Ofoq Normahmedov}
\author[a]{Jingzhe~Yang}
\author[a, 1]{Zhi~Zeng\note{Corresponding author.}}
\author[a]{Qian~Yue}
\author[a,b]{Jianping~Cheng}
\author[a]{Junli~Li}
\affiliation[a]{Key Laboratory of Particle and Radiation Imaging (Ministry of Education) and Department of Engineering Physics, Tsinghua University, Beijing 100084}
\affiliation[b]{College of Nuclear Science and Technology, Beijing Normal University, Beijing 100875}
\emailAdd{zengzhi@mail.tsinghua.edu.cn}

\abstract{A Geant4-based simulation framework for rare event searching experiments with germanium detectors named SAGE is presented with details. 
It is designed for simulating, assessing background distribution, and investigating the response of the germanium detectors. 
The SAGE framework incorporates its experiment-specific geometries and custom attributes, including the event generators, physics lists and output format, to satisfy various simulation objectives.
Its docker image has been prepared for virtualizing and distributing the SAGE framework.
Deployment of a Geant4-based simulation will be convenient under this docker image.
The implemented geometries include p-type point contact and broad energy germanium detectors with environmental surroundings, and these hierarchical geometries can be easily extended.
Users select these custom attributes via the JSON configuration files. 
The aforementioned attributes satisfy the simulation demands, and make SAGE a generic and powerful simulation framework for the CDEX experiment.
}


\begin{document}
\maketitle
\flushbottom

\section{Introduction}
The background models are essential for conceptual or technical design and data analysis of rare event searching experiments.
The background simulation and assessment raises the substantive requirements for the experimental design and material selections, and ultimately promotes the construction of the experimental setup \cite{Ma2019, Aprile2017}. 
While in the statistical analysis, a more precise background model ascertains a more convincing and robust conclusion\cite{Agostini2020, Cuesta2015, Aprile2019, Agnese2015, Kudryavtsev2015}. 
For rare event searching experiments, it is necessary to have a unified and convincing simulation framework for establishing their background models.
The simulation works can be conducted more efficiently within collaboration under these unified simulation frameworks \cite{Boswell2011}.
\par
SAGE (abbreviation for Simulation and Analysis for Germanium Experiments) is based on the Geant4 toolkit \cite{Agostinelli2003, Allison2006, Allison2016} and mainly developed by the China Dark matter EXperiment (CDEX) \cite{Jiang2018, Yang2019, Liu2019, She2020}, which is located in hall C of the China Jinping Laboratory phase-II (CJPL-II) \cite{Cheng2017}.
It is designed for simulating the background contributions for the ultra-low background germanium experiments such as dark matter (DM) searching, neutrinoless double beta decay (${0\nu \beta \beta}$) of ${^{76}}$Ge, as well as material screening by germanium gamma-ray spectrometers.
Currently, SAGE offers flexible interfaces for different experimental configurations.
It satisfies the need to accelerate the realizations and remove the challenges of simulations, such as knowledge of C++ and achieves consensus on the preset parameters in Geant4, so that the simulation works can become efficient, robust and reproducible.
SAGE allows its users to run a variety of simulations without writing any C++ code.
Its user-friendly functionalities and reliable preset parameters in the simulation workflow make SAGE a generic, powerful and user-friendly simulation framework.
\par
This paper describes the software framework and custom features of SAGE.
Its hierarchy and the detailed structures are discussed in Section~\ref{structure}. 
A configuration file is provided in Section~\ref{usecase}, presenting as a manual of SAGE. 
The last section summarizes all features and use of SAGE. 

\section{Structure of SAGE}
\label{structure}
\begin{figure}[htbp]
	\centering
	\includegraphics[width=0.8\textwidth]{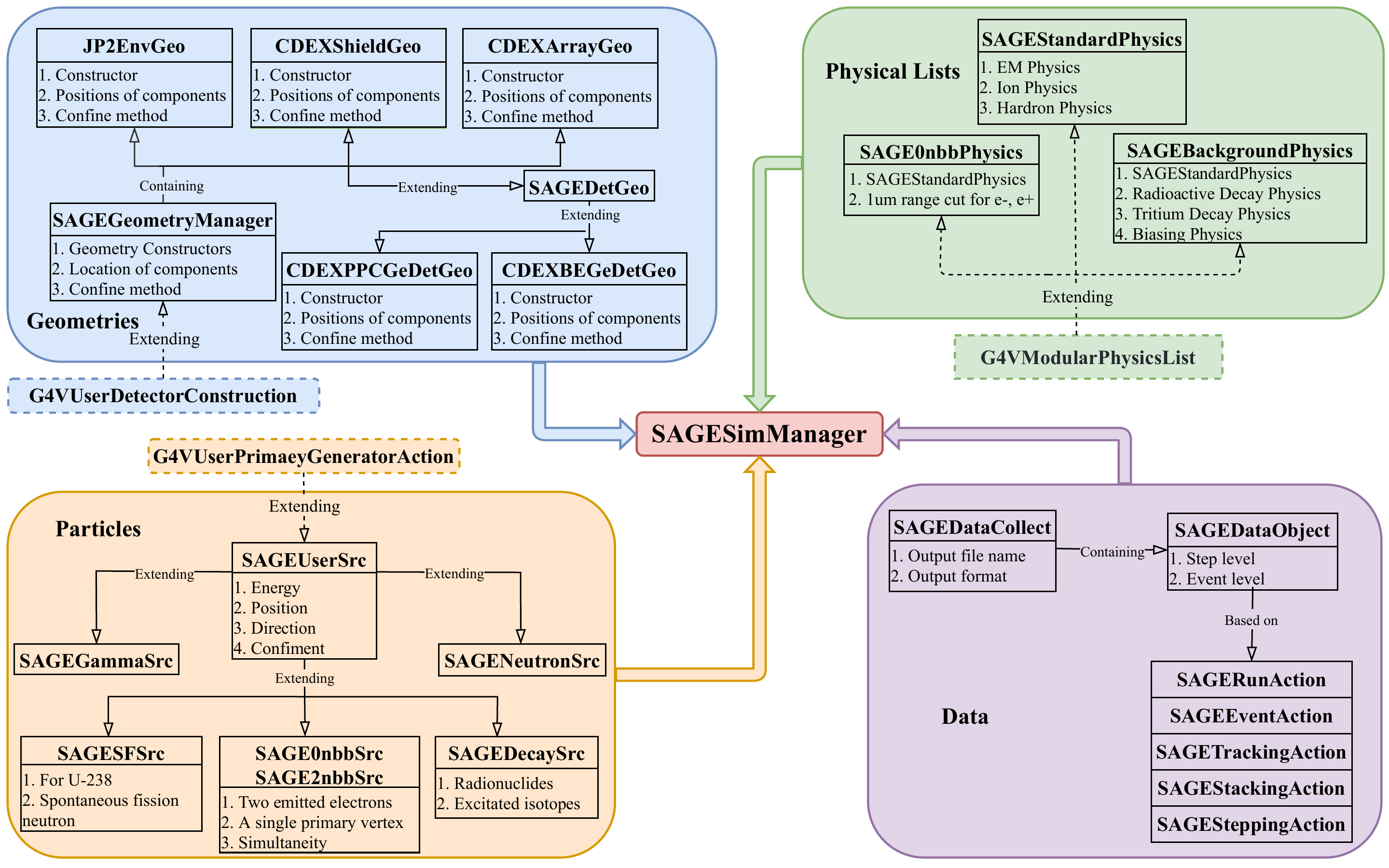}
    \caption{The software structure of SAGE follows the trace of Geant4 simulation. The JSON configuration file delivers the user-defined parameters to the SAGE simulation manager while this simulation manager controls of subsequent workflow.} 
	\label{Hierarchy}
\end{figure}

SAGE has been compiled and run with versions 10.6 and 10.7 of Geant4.
The modular design makes SAGE a flexible and extensible simulation framework.
Its hierarchy is depicted as shown in figure~\ref{Hierarchy}, and the following parts detail the implementation of each module.
The \textit{SAGESimManager} module supervises the simulation flow and enables the multi-threading capabilities provided by Geant4.
\subsection{Geometries}
This module is responsible for constructing the experimental geometries, and the Geant4 intrinsic geometry description classes support it.
SAGE provides a geometry manager for organizing hierarchical geometries and an abstract class \textit{SAGEDetGeo} for deriving different kinds of germanium detectors.
Custom geometries should be registered with the geometry manager.
\par
\begin{figure}[htbp]
	\begin{subfigure}{0.5\textwidth}
		\centering
		\includegraphics[width=0.8\linewidth]{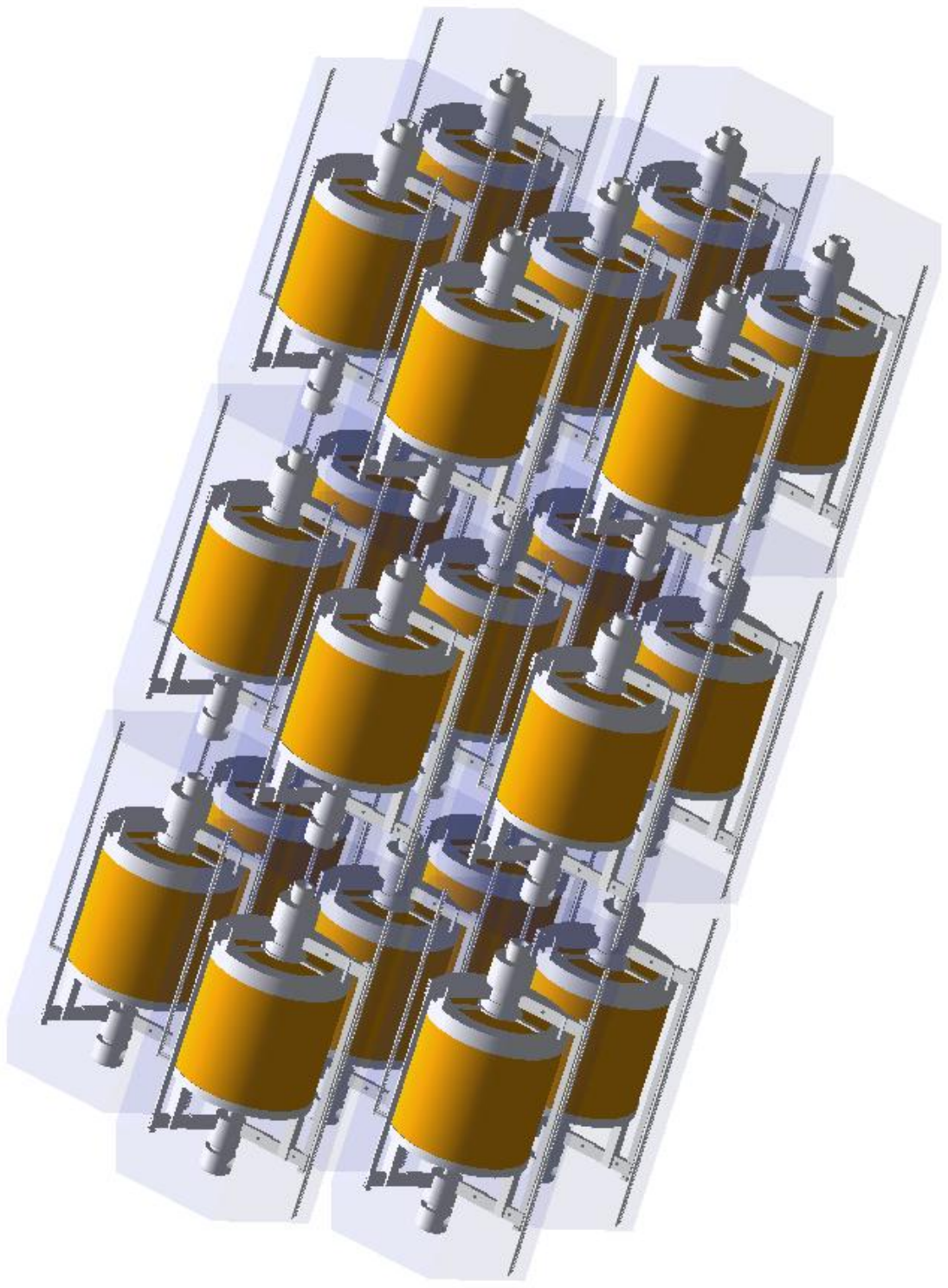}
	\end{subfigure}
	\begin{subfigure}{0.5\textwidth}
		\centering
		\includegraphics[width=0.4\linewidth]{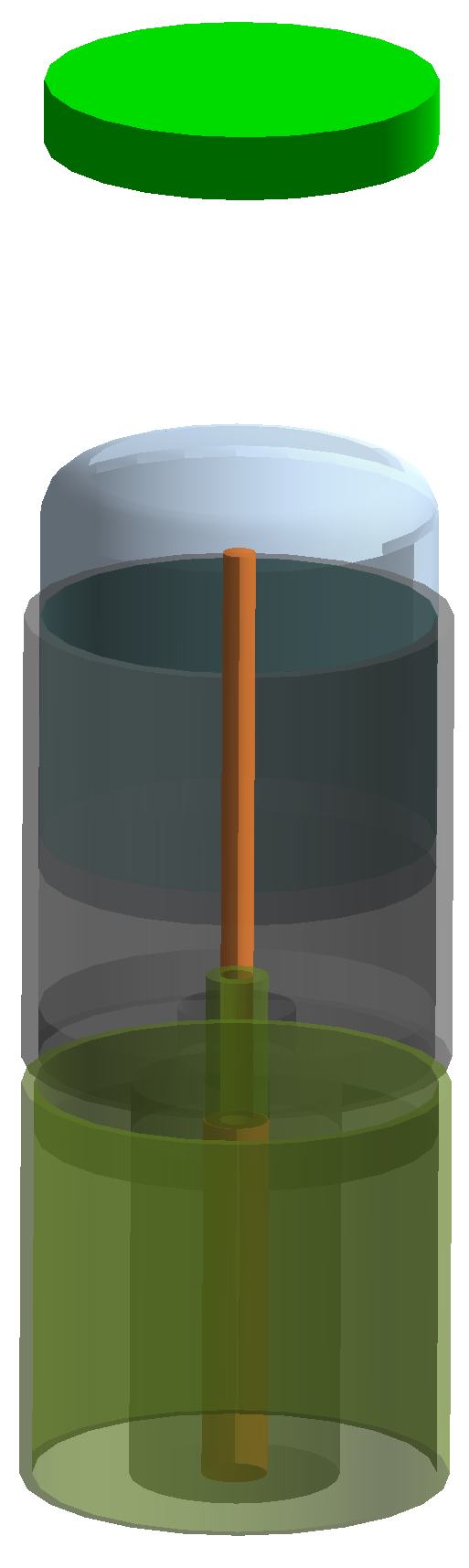}
	\end{subfigure}
    \caption{The geometric structures for germanium detectors of CDEX-100 (left) and GeTHU (right).}
    \label{germanium}
\end{figure}

Some experiments with germanium detectors are searching for rare events, including dark matter and ${0\nu \beta \beta}$ decay of $^{76}$Ge, such as SuperCDMS \cite{Amaral2020}, Gerda \cite{Agostini2020FinalResults}, CDEX \cite{Jiang2018}, EDELWEISS \cite{Arnaud2020}.
These experiments are conducted with the massive germanium detector as a detector array for enlarging their exposure.
Taken the CDEX experiment as an example, SAGE can be used for assessing the background distributions for these experiments.
Its germanium detector array is located in the stainless-steel cryostat filled with liquid nitrogen, as shown in figure~\ref{germanium}.
Therefore, this detector array can be dynamically enlarged, and the existing detector unit geometric classes can be reusable provided that they can be contained.
The germanium detectors are also highly used as gamma-ray spectrometers, such as GeTHU \cite{Zeng2014}, Gator \cite{Baudis2011}, BUGS \cite{Scovell2017}.
SAGE assists in the construction of their background models and the simulation of their detection efficiencies.
SAGE has been used in the simulation for GeTHU whose geometry is shown in figure~\ref{germanium}.

\subsection{Event generator}
Customized event generators originate from an abstract class \textit{SAGEUserSrc} which is based on the \textit{G4SingleParticleSource}.
These customized event generators produce particles including neutron, electron, gamma, radionuclides, ${0\nu \beta \beta}$ and ${2\nu \beta \beta}$ isotope source \cite{Tretyak1995}.
Although the ${0\nu \beta \beta}$ and ${2 \nu \beta \beta}$ sources cannot be generated with \textit{G4SingleParticleSource} directly, two electrons are attached to the primary vertex to generate two electrons simultaneously by an embedded algorithm.
In addition to the point, surface and volume sources, we attach the CDEX geometric volumes to these generators through a member function (named \textit{ConfinePosition}) of the abstract class \textit{SAGEUserSrc} to facilitate the sampling strategies.

\subsection{Physics Lists}
The physics list module is constructed and optimized for germanium detectors by capitalizing the Geant4 intrinsic physics lists.
It was designed according to the suggestions of the Geant4 team and the related examples in the Geant4 package.
Three available physics lists \textit{SAGEStandardPhysics}, \textit{SAGE0nbbPhysics} and \textit{SAGEBackgroundPhysics} are optimized for different simulation projects.
Among these physics lists, the \textit{SAGEStandardPhysics} is the simplest and most efficient one.
The \textit{SAGEStandardPhysics} physics list consists of modular physics lists including \textit{G4DecayPhysics, G4EmStandardPhysics\_option3, G4EmExtraPhysics, G4HadronPhysicsFTFP\_BERT, G4HadronElasticPhysics, G4StoppingPhysics, G4IonPhysics}.
These modular physics lists satisfy the simulation requirements of germanium detectors in most cases, and \textit{SAGEStandardPhysics} is used by default, but more requirements are raised in some instances.
Based on \textit{SAGEStandardPhysics}, the other two physics lists integrate more features for their simulation aims.
\textit{SAGE0nbbPhysics} is tailored for tracking electrons with a smaller step limit of electrons (1 ${\mu}$m rather than 1 mm default value) with about 15 times larger than CPU consuming time.
With an 1-${\mu}$m step limit for an electron, its track will be terminated when its energy is less than 10 keV, but an 1-mm step limit terminates for 800 keV. 
It can be applied to investigate the characteristics of pulses generated from the charge collections or the pre-amplifier.
\textit{SAGEBackgroundPhysics} is intended for the ambient and other intrinsic background contributions of germanium detectors.
\textit{SAGEBackgroundPhysics} can be used to establish background models and assess the background indices for experiments.
\textit{SAGEBackgroundPhysics} has two additional features.
\par
\begin{enumerate}[label=(\arabic*), start = 1]
    \item The ${\beta}$-decay of ${^3}$H.
    \par
    ${^3}$H is the product of cosmogenic activation inside the germanium crystal, decays and then emits an electron with a maximum energy of 18.6 keV and an anti-neutrino.
    It is treated as a stable isotope in Geant4, however.
    Users must calculate the energy spectrum of ${^3}$H ${\beta-}$decay before simulations and sample the incident electrons in deference to their energy distribution.
    SAGE instantiates a relevant physics sub-list named \textit{SAGETritiumPhysics} to make ${^3}$H decay automatically in Geant4.
\par
    \item Biasing sampling
    \par
    The second one is a generic biasing physics sub-list.
    When a radioactive source locates far from the detector (a remote radioactive source) or the cross-sections of certain interactions are too small to obtain the statistically significant results, such as simulating the gamma-rays background contribution decreased by shielding, simulation with ${2^{31}}$ incident particles (the maximum value of a single run in Geant4) probably cannot get an acceptable result unless simulating multiple runs.
    However, multiple runs will consume more computation time, and the results may still not be meaningful within reasonable simulation events under certain circumstances.
    SAGE has integrated two kinds of biasing techniques inside this generic biasing sub-list to alleviate these dilemmas.
    Russian roulette and split strategy \cite{MCBook}, commonly used in deep penetration problems, is embedded in SAGE to tackle the simulation issue when the remote sources are distributed outside shielding. 
    This method is similar to the Geant4 example (\textit{extended/biasing/GB03}) other than the necessary concatenations with other SAGE modules.
    The shielding should be divided into multiple layers depending on their locations, and particles will participate in the biasing games when they occur at the boundaries of layers.
    When a particle in the direction of detectors occurs at the boundaries of layers, its track will be cloned multiple times.
    And then, these tracks will be simulated separately with the weight divided by the number of tracks.
    Otherwise, it might be killed according to the surviving probability ${\mathcal{P}}$ defined by users.
    If a particle survives Russian roulette, its track will have a weight increased by a factor of ${1/\mathcal{P}}$.  
    More computation power is assigned to particles with greater possibilities of reaching the germanium detectors through the split and kill method.
	When considering the interactions with tiny cross-sections such as neutron originated from ${(\alpha, n)}$ interactions, their cross-sections should be artificially increased before simulation; otherwise, they will not be observed.
	Although a tiny cross-section weakens the potential for neutron ${(\alpha, n)}$ interactions, the enormous mass of surrounding construction materials, such as concrete or stainless steel cryostat, makes neutrons detectable in the ultra-low background experiments.
    To make them also observable in the simulations within limited events, SAGE adopts the change cross-section biasing method \cite{MCBook2} as a variant of the strategy described in \textit{extended/biasing/GB01}.
    When calculating the mean free path, the contributions from selected interactions are multiplied by a transformation factor determined by users.
    Then more events interact via the selected interactions, and more computation power is assigned to these interactions.
    Further discussion on this method can be found in \cite{Mendoza2020}.
\end{enumerate}

\subsection{Output}
SAGE instantiates a \textit{SAGEDataCollect} object to gather the interaction and trajectory information as particles interact with materials step by step. 
SAGE provides two intrinsic analysis levels with the \textit{SAGEDataObject} object.
The former one is called the "event" level, which analyzes all steps in a single event to get its energy deposition (grouped by the interaction mediators, such as electrons, neutrons, other particles), track length, living time(the time since it is created) and weight.

In simulations of radioactive decay chains, SAGE adopts the separation time of 10 ${\mu s}$, roughly the detector response time, to distinguish radiogenic ${\alpha, \beta}$ and ${\gamma}$ rays from different isotopes in a single radioactive chain while their energies are by default recorded in a single event by Geant4. 
If the global time of the current step is greater than 10 ${\mu s}$, the simulation manager will reset its global time and record all interactions before the current step.
The latter one named "step" level, recording and dumping all step information for debugging or pulse shape discrimination, contains the PDG encoding of particles, energy deposition, kinetic energy, interaction length, living time, weight, velocity direction, position and weight.
Both levels have three output formats, i.e., *.csv, *.root or *.h5, for further analysis with the specified analysis tools.

\section{Use Case}
\label{usecase}
Users can adjust the free parameters in the configuration files.
SAGE is assisted by a third-party module of JSON parser \cite{JsonParser} to convert the configuration files into a dictionary data structure. 
A set of applicable JSON configuration files are provided in the SAGE software package as the user guides.
The configuration files are arranged in six parts concerning the free simulation ingredients as follows:
\begin{enumerate}[label = (\arabic*), start = 1]
\item Mode: Two modes are provided. 
    The "Simulation" mode means conducting simulation more efficiently with command lines, while the "Visualization" mode is helpful when you are curious about or debugging the simulation geometries.
    The visualization mode depends on the Geant4 visualization setup and is also named interactive mode due to its instant response to the manual input messenger command.  
\item Physics: There exist three physics lists related to different simulation goals. 
    The "background" physics list contains the radioactive decay processes while the "0nbb" physics list has a smaller step limit for electrons. 
    The "standard" physics list can be used if time efficiency is sensitive.
\item Array: Constructional parameters of the germanium detector arrays, containing the number of detectors, the distance between detectors, the type and material ("natural" or "enriched" germanium) of germanium detectors.
\item Particle: The information of incident particles, such as particle name, energy, momentum direction, position and the number of incident particles.
\item Biasing: Status and parameters of biasing techniques. In a normal case, biasing is disabled by default. 
\item Output: The output file name and the output level ("event" or "step" data object).
\end{enumerate}
All of these properties have their default values for removing redundant parameters.
For example, the biasing methods are disabled in default, so they can be neglected if not necessary. 
\par
\begin{figure}[H]
	\centering
	\includegraphics[width=0.6\textwidth]{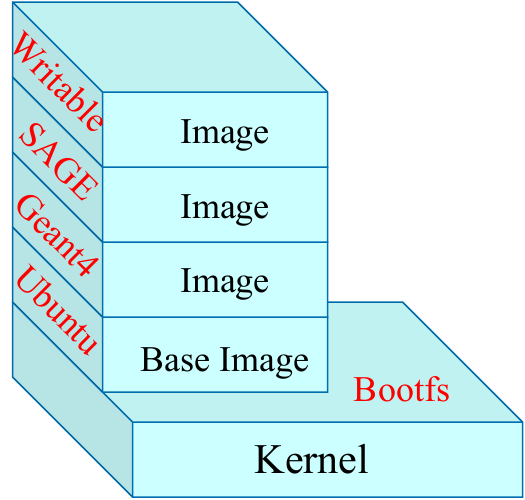}
	\caption{Docker image for SAGE.}
	\label{Docker}
\end{figure}

SAGE has been compiled and linked into a single executable application together with a set of instructive configuration files.
To distribute it more conveniently, SAGE and Geant4 have been wrapped inside a single docker image.
The docker container compiles the Geant4.10.7 and SAGE, so users run SAGE directly after pulling.
With a docker container, it is much easier for users to customize, distribute and use SAGE. 
The structure of this container is shown in figure~\ref{Docker}.
In addition, the docker file is also provided to build images for users' demands.  

\par
\begin{figure}[htbp]
    \centering
    \includegraphics[width=0.48\textwidth]{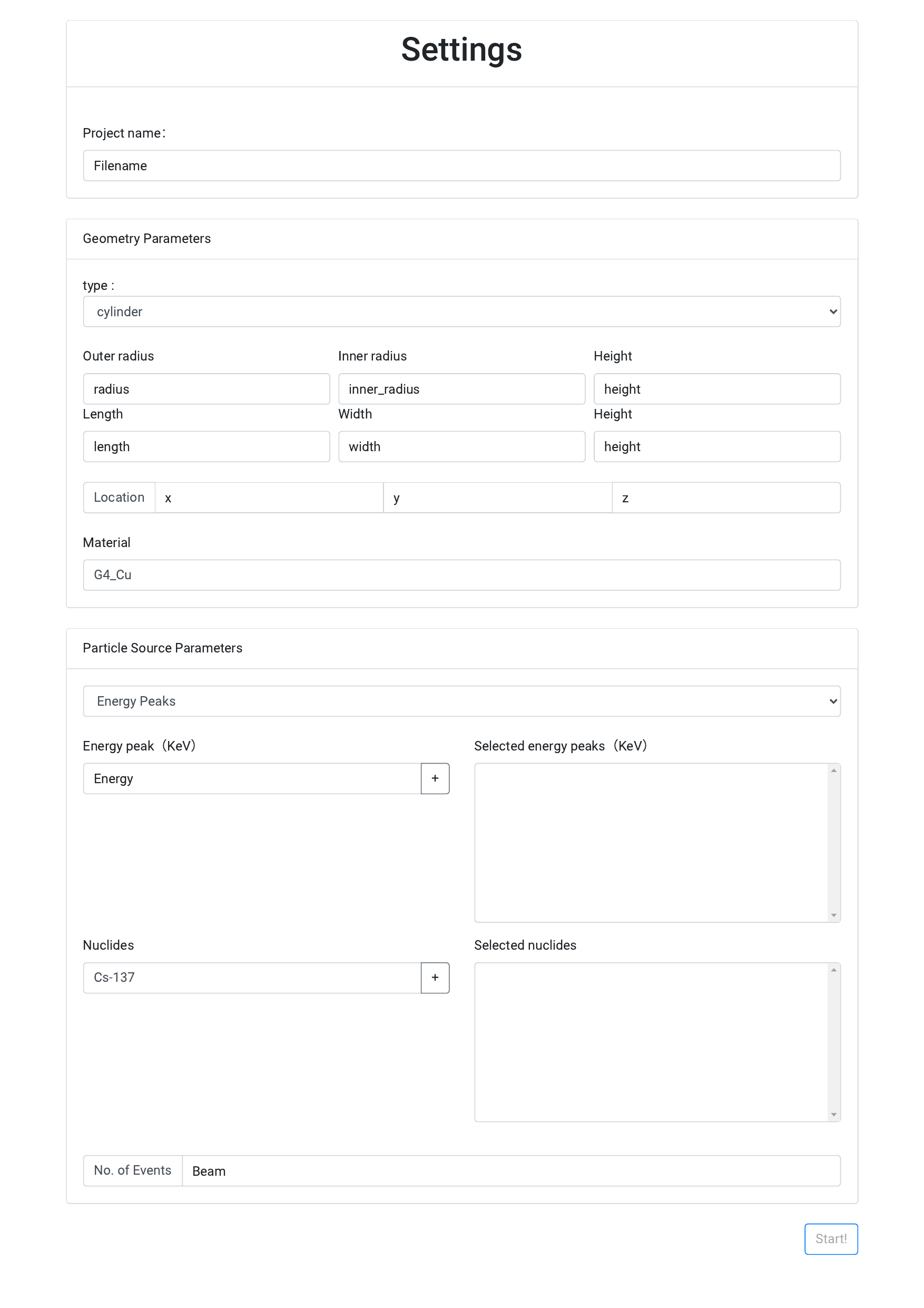}
    \caption{The default GUI of SAGE designed for GeTHU gamma-ray spectrometer.}
    \label{GUI}
\end{figure}

A graphical user interface (GUI) with Electron, a framework for creating native applications with web technologies, is developed to make it more user-friendly to newbies.
The GUI takes charge of generating configuration files and provides a shortcut to start the simulation task with a single click.
The parameters shown in the GUI can be modified to satisfy the keywords defined by users.
For GeTHU, Its GUI is shown in figure~\ref{GUI}.
The interface is divided into three categories.
The file name should be written in the card named \textit{Project name}.
The geometry of samples can be selected in the card named \textit{Geometry Parameters} which is consisted of several arguments including shape (cylinder or cuboid), dimensions of shapes, location and materials.
The energy regions or nuclides of interest are added and stored with a list shown on the right via the card named \textit{Particle Source Parameters}.
The number of simulated events is also included in this card.
SAGE will be executed with the configuration generated by this GUI tool by clicking the start button at the right bottom corner.
\par

\begin{figure}
	\centering
	\includegraphics[width=0.85\textwidth]{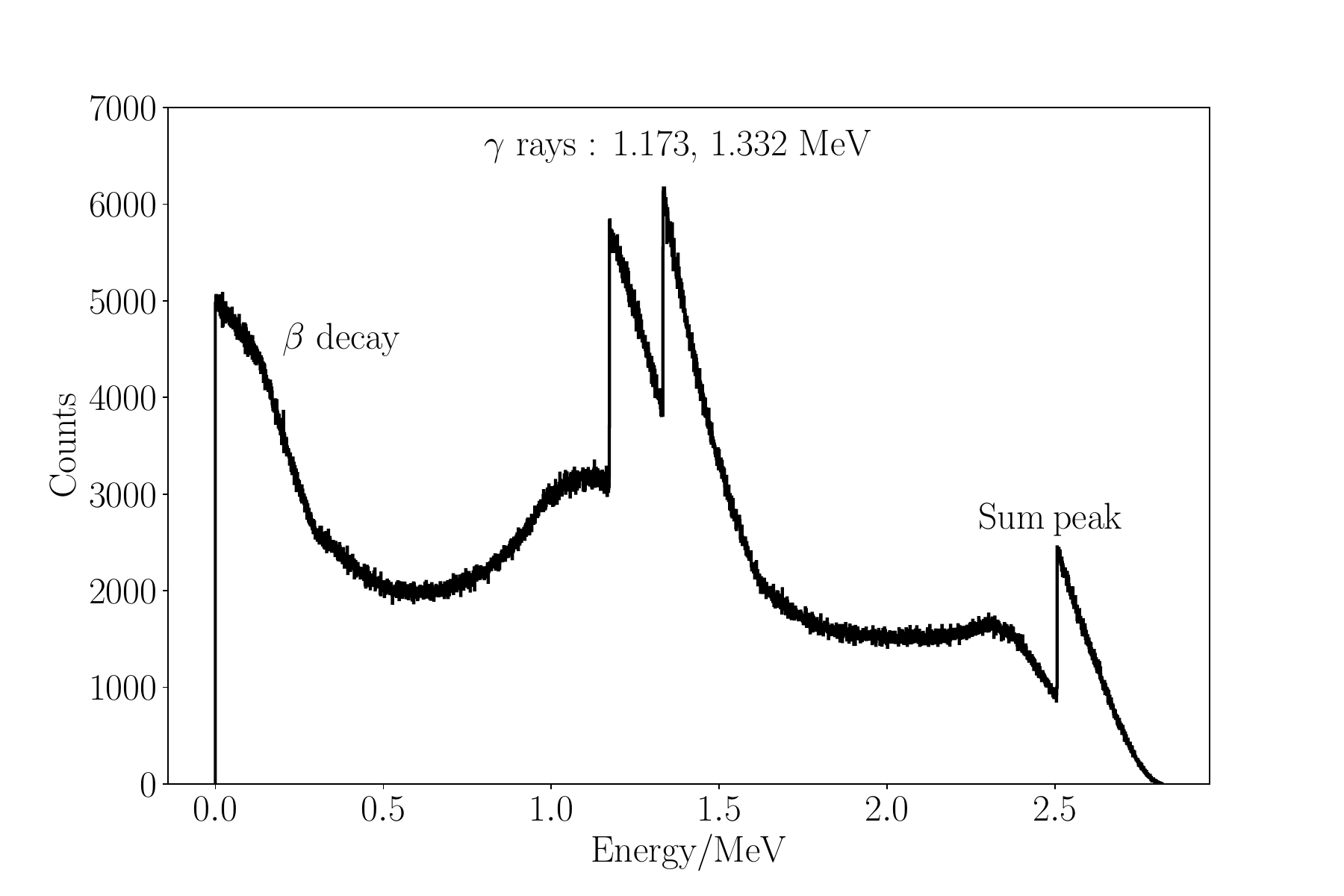}
	\caption{The simulated energy spectrum of cosmogenic ${^{60}}$Co in germanium crystal.}
	\label{Co60Ge}
\end{figure}

SAGE has been widely used to simulate energy spectra in germanium detectors induced by radioactive samples and background sources, e. g. cosmogenic and primordial radionuclides for GeTHU \cite{Zeng2014} and CDEX experiment \cite{Ma2020}.
For example, the radionuclides ${^{60}}$Co in the germanium crystal for the CDEX-100 experiment can be simulated with the configuration below:
\begin{lstlisting}[language=C]
{
  "Mode" : "Simulation",
  "Physics" : "Background",
  "Array" :
  {
    "HeightGap" : 0,
    "RadiusGap" : 0,
    "HeightN" : 5,
    "RadiusN" : 5,
    "Material" : "enriched",
    "Type" : "BEGe"
  },
  "Particle" :
  {
    "Type" : "Decay",
    "Distribution" : "Component",
    "Confine" : "physGe",
    "Isotope" : "Co-60",
    "Beam" : 10000000
  },
  "Biasing" :
  {
    "GeoBias" : "Off",
  },
  "Output" :
  {
    "Filename" : "Co60-Ge",
    "DataMode" : "event"
  }
}
\end{lstlisting}
When we execute the SAGE with the configuration file above, we will get all simulation events in the Co60-Ge\_nt\_data.csv. 
Based on this output, the expected energy spectrum can be reconstructed, as shown in figure~\ref{Co60Ge}. 
Three peaks shown in figure~\ref{Co60Ge} are two gamma-ray peaks (1.173 and 1.332 MeV) of ${^{60}}$Co and the sum peak of these two gamma rays, respectively.
Besides, the Compton edges of two gamma-ray peaks lead to an unusual increasing between 0.6 and 1.1 MeV energy region. 
In the case of intrinsic background simulation inside the germanium crystal, the contributions of ${\beta}$-decay are also visible in the low-energy region of the energy spectrum.
The events recording the energy deposition of both ${\beta}$-decay and subsequent gamma rays contribute to the non-gaussian shape of the gamma-ray peaks.
\par
SAGE is validated by the calculations on the cross-sections of proton activations.
Five natural lead samples were irradiated by different energies (40, 70, 100, 400 and 800 MeV) protons from 2000 to 2002 \cite{Titarenko2006} and were measured with GeTHU in September 2020 at CJPL.
SAGE is used to simulate their energy spectra of proton-induced radioactive isotopes, and then the measured energy spectrum is decomposed into these simulated energy spectra.
The cross-sections are calculated based on the decomposed spectra and compatible with previous results \cite{Titarenko2020}.
\par

Thanks to its modular design, SAGE can be widely used and extended in various experiments with different physics goals.
For instance, the background spectra can be simulated after implementing the experiment-specific geometries inherited from the abstract class \textit{SAGEDetGeo}.
It assists the experimental setup design, assesses the background level, investigates detector response and estimates the projected sensitivities for experiments with germanium detectors.  

\section{Summary}
In this work, we describe the simulation framework of SAGE and its customized features to simulate the particle trajectories in experiments using germanium detectors.
Its configuration files are written in JSON format, and flexible parameters in the simulation workflow can be determined.
SAGE has integrated the physics list for ${^3}$H decay and biasing sampling techniques.
All customized attributes make SAGE a unified, user-friendly, robust and reproducible simulation framework for germanium experiments and SAGE widely used in the CDEX collaboration.
Currently, additional features are under investigation as follows to broaden its application.
\begin{itemize}
    \item More geometries  
    \item Statistical methods for further data analysis
    \item Pulse shape simulations and discriminations
\end{itemize}
Although SAGE is designed for experiments using germanium detectors, its implemented modules, except experiment-specific geometries, can be widely used in other rare event searching experiments having different detection techniques, which makes SAGE a more powerful and general simulation framework.
\acknowledgments
This work was supported by the National Key Research and Development Program of China (Grant No.2017YFA0402201), the National Natural Science Foundation of China (Grants No.U1865205 \& 11725522 \& 11675088), and Tsinghua University Initiative Scientific Research Program (No.2019050007).

\bibliographystyle{JHEP}
\bibliography{SAGE}

\providecommand{\href}[2]{#2}\begingroup\raggedright\begin{thebibliography}{10}

\bibitem{Ma2019}
J.~L. Ma et~al., \emph{{Study on Cosmogenic Activation in Germanium Detectors
  for Future Tonne-scale CDEX Experiment}}, {\emph{Sci. China Phys. Mech.}
  {\bfseries 62} (2019) 11011}.

\bibitem{Aprile2017}
E.~Aprile et~al., \emph{{Material Radioassay and Selection for the XENON1T Dark
  Matter Experiment}}, {\emph{Eur. Phys. J. C} {\bfseries 77} (2017) 890}.

\bibitem{Agostini2020}
M.~Agostini et~al., \emph{{Modeling of Gerda Phase II data}}, {\emph{J. High
  Energy Phys.} {\bfseries 3} (2020) 139}.

\bibitem{Cuesta2015}
C.~Cuesta et~al., \emph{{Background Model for the Majorana Demonstrator}},
  {\emph{Phys. Proc.} {\bfseries 61} (2015) 821}.

\bibitem{Aprile2019}
{\scshape XENON} collaboration, \emph{{XENON1T Dark Matter Data Analysis:
  Signal Reconstruction, Calibration, and Event Selection}}, {\emph{Phys. Rev.
  D} {\bfseries 100} (2019) 052014}.

\bibitem{Agnese2015}
R.~Agnese, \emph{{Maximum Likelihood Analysis of Low Energy CDMS II Germanium
  Data}}, {\emph{Phys. Rev. D} {\bfseries 91} (2015) 052021}.

\bibitem{Kudryavtsev2015}
V.~A. Kudryavtsev and J.~L. Orrell, \emph{{Expected Background in the LZ
  Experiment}}, {\emph{AIP Conf. Proc.} {\bfseries 1672} (2015) 060003}.

\bibitem{Boswell2011}
M.~Boswell et~al., \emph{{MaGe-a Geant4-Based Monte Carlo Application Framework
  for Low-Background Germanium Experiments}}, {\emph{IEEE T. Nucl. Sci.}
  {\bfseries 58} (2011) 1212}.

\bibitem{Agostinelli2003}
S.~Agostinelli et~al., \emph{{Geant4—A Simulation Toolkit}}, {\emph{Nucl.
  Instrum. Meth. A} {\bfseries 506} (2003) 250}.

\bibitem{Allison2006}
J.~Allison et~al., \emph{{Geant4 Developments and Applications}}, {\emph{IEEE
  T. Nucl. Sci.} {\bfseries 53} (2006) 270}.

\bibitem{Allison2016}
J.~Allison et~al., \emph{{Recent Developments in Geant4}}, {\emph{Nucl.
  Instrum. Meth. A} {\bfseries 835} (2016) 186}.

\bibitem{Jiang2018}
{\scshape CDEX} collaboration, \emph{{Limits on Light Weakly Interacting
  Massive Particles from the First ${102.8\text{kg}\times\text{day}}$ Data of
  the CDEX-10 Experiment}}, {\emph{Phys. Rev. Lett.} {\bfseries 120} (2018)
  241301}.

\bibitem{Yang2019}
{\scshape CDEX} collaboration, \emph{{Search for Light
  Weakly-Interacting-Massive-Particle Dark Matter by Annual Modulation Analysis
  with a Point-Contact Germanium Detector at the China Jinping Underground
  Laboratory}}, {\emph{Phys. Rev. Lett.} {\bfseries 123} (2019) 221301}.

\bibitem{Liu2019}
{\scshape CDEX} collaboration, \emph{{Constraints on Spin-Independent Nucleus
  Scattering with sub-GeV Weakly Interacting Massive Particle Dark Matter from
  the {CDEX-1B} Experiment at the China Jinping Underground Laboratory}},
  {\emph{Phys. Rev. Lett.} {\bfseries 123} (2019) 161301}.

\bibitem{She2020}
{\scshape CDEX} collaboration, \emph{{Direct Detection Constraints on Dark
  Photons with the CDEX-10 Experiment at the China Jinping Underground
  Laboratory}}, {\emph{Phys. Rev. Lett.} {\bfseries 124} (2020) 111301}.

\bibitem{Cheng2017}
J.~P. Cheng et~al., \emph{{The China Jinping Underground Laboratory and Its
  Early Science}}, {\emph{Annu. Rev. Nucl. Part. Sci.} {\bfseries 67} (2017)
  231}.

\bibitem{Amaral2020}
D.~W. Amaral et~al., \emph{{Constraints on low-mass, relic dark matter
  candidates from a surface-operated SuperCDMS single-charge sensitive
  detector}}, {\emph{Phys. Rev. D} {\bfseries 102} (2020) 091101}.

\bibitem{Agostini2020FinalResults}
{\scshape GERDA} collaboration, \emph{{Final Results of GERDA on the Search for
  Neutrinoless Double-$\ensuremath{\beta}$ Decay}}, {\emph{Phys. Rev. Lett.}
  {\bfseries 125} (2020) 252502}.

\bibitem{Arnaud2020}
{\scshape EDELWEISS} collaboration, \emph{{First Germanium-Based Constraints on
  Sub-MeV Dark Matter with the EDELWEISS Experiment}}, {\emph{Phys. Rev. Lett.}
  {\bfseries 125} (2020) 141301}.

\bibitem{Zeng2014}
Z.~Zeng et~al., \emph{{The Characteristics of a Low Background Germanium Gamma
  Ray Spectrometer at China JinPing underground laboratory}}, {\emph{Appl.
  Radiat. Isotopes} {\bfseries 91} (2014) 165}.

\bibitem{Baudis2011}
L.~Baudis, A.~D. Ferella, A.~Askin et~al., \emph{{Gator: a low-background
  counting facility at the Gran Sasso Underground Laboratory}}, {\emph{J.
  Instrum.} {\bfseries 6} (2011) P08010}.

\bibitem{Scovell2017}
P.~R. Scovell, E.~Meehan, H.~M. Araújo et~al., \emph{{Low-background gamma
  spectroscopy at the boulby underground laboratory}}, {\emph{Astropart. Phys.}
  {\bfseries 97} (2017) 160}.

\bibitem{Tretyak1995}
V.~I. Tretyak and Y.~G. Zdesenko, \emph{{Tables of Double Beta Decay Data}},
  {\emph{Atom. Data Nucl. Data} {\bfseries 61} (1995) 43}.

\bibitem{MCBook}
\emph{{Monte Carlo integrations II: Improving efficiency}},  in
  \emph{{Physically Based Rendering}}, M.~Pharr and G.~Humphreys, eds.,
  (Burlington), pp.~663 -- 718, Morgan Kaufmann, (2004).

\bibitem{MCBook2}
C.~P. Robert and G.~Casella, \emph{{Monte Carlo Integration}},  in \emph{{Monte
  Carlo Statistical Methods}}, (New York, NY), pp.~79--122, Springer New York,
  (2004).

\bibitem{Mendoza2020}
E.~Mendoza et~al., \emph{{Neutron Production Induced by ${\alpha}$-Decay with
  Geant4}}, {\emph{Nucl. Instrum. Meth.} {\bfseries 960} (2020) 163659}.

\bibitem{JsonParser}
``nlohmann/json.'' \url{http://github.com/nlohmann/json}.

\bibitem{Ma2020}
H.~Ma et~al., \emph{{In-situ Gamma-ray Background Measurements for Next
  Generation CDEX Experiment in the China Jinping Underground Laboratory}},
  {\emph{Astropart. Phys.} {\bfseries 128} (2021) 102560}.

\bibitem{Titarenko2006}
Y.~E. Titarenko et~al., \emph{{Excitation functions of product nuclei from 40
  to 2600MeV proton-irradiated 206,207,208,natPb and 209Bi}}, {\emph{Nucl.
  Instrum. Meth. A} {\bfseries 562} (2006) 801}.

\bibitem{Titarenko2020}
Y.~Titarenko et~al., \emph{{$^{208,207,206,nat}$Pb(p,x)$^{207}$Bi and
  $^{209}$Bi (p,x)$^{207}$Bi excitation functions in the energy range of 0.04 -
  2.6 GeV}},
  \href{https://doi.org/https://doi.org/10.1016/j.nima.2020.164635}{\emph{Nuclear
  Instruments and Methods in Physics Research Section A: Accelerators,
  Spectrometers, Detectors and Associated Equipment} {\bfseries 984} (2020)
  164635}.

\end{thebibliography}\endgroup
\end{document}